\documentclass[conference]{IEEEtran}
\usepackage{amsmath,amsfonts}
\usepackage{algorithmic}
\usepackage{algorithm}
\usepackage{array}
\usepackage[caption=false,font=normalsize,labelfont=sf,textfont=sf]{subfig}
\usepackage{textcomp}
\usepackage{stfloats}
\usepackage{subfig}
\usepackage{url}
\usepackage{verbatim}
\usepackage{graphicx}
\usepackage{cite}
\usepackage{amsmath}
\begin{document}

\title{Complexity and Coupling: A Functional Domain Approach}

\author{\IEEEauthorblockN{Aydin Homay}
\IEEEauthorblockA{\textit{Chair of Industrial Communications} \\
\textit{Technische Universität Dresden)}\\
Dresden, Germany \\
https://orcid.org/0000-0002-6425-7468}
}

\markboth{Journal of \LaTeX\ Class Files,~Vol.~14, No.~8, August~2021}%
{Shell \MakeLowercase{\textit{et al.}}: A Sample Article Using IEEEtran.cls for IEEE Journals}


\maketitle

\begin{abstract}
This paper provides a precise and scientific definition of complexity and coupling, grounded in the functional domain, particularly within industrial control and automation systems (iCAS). We highlight the widespread ambiguity in defining complexity and coupling, emphasizing that many existing definitions rooted in physical attributes lead to confusion and inconsistencies. Furthermore, we re-exhibit why coupled design inherently increases complexity and how potentially this complexity could be reduced. Drawing on examples from various disciplines, such as software engineering, industrial automation, and mechanical design, we demonstrate that complexity does not necessarily correlate with system size or the number of components, and coupling, unlike common belief in software engineering, actually does not occur in the physical domain but in the functional domain. We conclude that effective design necessitates addressing coupling and complexity within the functional domain.
\end{abstract}


\section{Introduction}
\IEEEPARstart{T}o understand why a coupled design is complex, we must first establish clear and scientific definitions of both coupling and complexity. Unfortunately, the software industry lacks such definitions, which impedes progress in these areas. As Suh notes, meaningful advancement is impossible when a concept is interpreted differently depending on whom we ask~\cite{Suh-2005}. Thus, advancing any field requires precise and shared definitions.

Before we define complexity and coupling scientifically, we must first clarify what constitutes a scientific definition and how it differs from other forms of definition. Fiona J. Hibberd defines a scientific definition as one that identifies the essential features of a concept—features without which the concept would not be what it is. Grounded in realism, scientific definitions follow a genus-differentia structure: they classify an entity (genus) and specify its distinguishing properties (differentia). For instance, water is defined as a substance (genus) characterized by the molecular structure H\textsubscript{2}O (differentia)~\cite{Hibberd-2019}.

Similarly, the first and second laws of thermodynamics serve as examples of scientific definitions. They define fundamental properties of energy and entropy that are invariant across systems and independent of specific physical implementations. The first law—conservation of energy—defines energy as a measurable quantity that cannot be created or destroyed, only transformed. The second law introduces the concept of entropy and defines the directionality of energy transformations. Both laws identify essential properties and remain valid regardless of the particular machine, substance, or context involved. Their universality and independence from specific implementations demonstrate the characteristics of scientific definitions.

In this sense, Nam P. Suh is correct to argue against defining coupling and complexity solely within the physical domain (e.g., machines, lines of code, or computation time). Physical definitions hinder the formation of a stable genus-differentia structure, as the choice of genus (e.g., machine or code) inevitably alters the differentia~\cite{Suh-2005}. Therefore, two conditions must be met when defining coupling and complexity: (1) the definition must remain consistent regardless of who formulates it, and (2) it must identify essential, distinguishing properties.

The remainder of the paper is structured as follows: Section~\ref{ambiguity_of_complexity} analyzes the ambiguity surrounding the concept of complexity and reviews domain-specific interpretations. Section~\ref{defination_of_conplexity} introduces a scientific definition of complexity rooted in the functional domain. Section~\ref{Formulation_of_complexity} presents a mathematical formulation of complexity using information theory. Section~\ref{complexity_in_software_systems} examines how complexity has historically been defined in software systems. Section~\ref{what_is_coupling} provides a detailed discussion of coupling, contrasting physical and functional domain definitions. Section~\ref{running_example} introduces a running example in the context of industrial automation to ground the theory in practice. Section~\ref{why_compled_design_is_complex} explains why coupled design inherently leads to increased complexity. Finally, Section~\ref{conclusion} concludes the paper and outlines future research directions.

\section{Ambiguity of Complexity}
\label{ambiguity_of_complexity}
Nam P. Suh in his book "Complexity: Theory and Applications" argues that the majority of individuals appear to have an intuitive grasp of what complexity implies. However, when we dig deeper into their perceptions and comprehension, we uncover a myriad of distinct perspectives on the concept. This diversity of interpretation is evident even among students and academics who specialize in the study of complexity. These individuals often use the term complexity to articulate a variety of notions and viewpoints. He continues his argument that if we do not define complexity in specific terms, we cannot advance it because depending on whom we ask, we will get a different definition of complexity~\cite[pp.~4]{Suh-2005}. 

In the disciplines of mathematics and computer science, there has been ongoing debate regarding the quantification of the complexity associated with algorithms, stochastic processes, material structures, and computational processes. Feldman and Crutchfield~\cite{Feldman-Crutchfield-1998,Cormen-2022} provide a comprehensive examination of various "complexity measures" in these domains, including Kolmogorov-Chaitin complexity, computational complexity, stochastic complexity, statistical complexity, and structural complexity. However, it appears that not all these complexity measures are effective in facilitating the rational and simplified design of engineering systems, nor in reducing their complexity, since these metrics are predominantly grounded in the physical domain rather than the functional domain and are confined to specific contexts. The definition of complexity posited in this paper aims to address these concerns~\cite{Suh-Complexity-2005}.

In the discipline of engineering, the ultimate goal is to reduce the complexity of engineered systems by employing a rational design methodology grounded in fundamental principles. This is aimed at enhancing the systems’ reliability, reducing the cost of development and operation, and enhancing their performance. In dealing with complexity in natural science, the objective is to comprehend the complexity of nature to forecast the behavior of natural systems and to leverage this fundamental knowledge to address complex issues within biology, physics, and information sciences. In the realm of social science, the goal is to understand the complexity issues related to the causality of societal problems and ultimately, design solutions to these societal problems that have minimal complexity~\cite{Suh-Complexity-2005}. But this can only be achieved if we first give a scientific definition to what complexity is and how to understand it~\cite{Suh-Complexity-2005,Suh-2005}; without such a foundation, we either risk creating a tangle of incompatible definitions or falling into human cognitive traps (e.g., cognitive bias, hallucination)~\cite{Snowden-2024}.

\section{Defination of Conplexity}
\label{defination_of_conplexity}
As stated earlier, defining complexity in the physical domain (e.g., machines, lines of code, computation time) can lead to ambiguous and inconsistent interpretations. On the other hand, the complexity defined in the functional domain is a measure of uncertainty in achieving a set of tasks defined by functional requirements~\footnote{Functional requirements are by definition a minimum set of independent requirements~\cite[pp.~14]{Suh-2001}.} in the functional domain. The "functional" approach is to treat complexity as a relative concept that evaluates how well we can satisfy "what we want to achieve" with "what is achievable". This approach requires a mapping from the "functional domain" to "physical domain" to determine the complexity. The complexity based on physical thinking would have field-specific "dimensions," whereas the complexity based on functional thinking would be dimensionless regardless of the specific subject under consideration~\cite{Suh-Complexity-2005}.

Therefore, complexity should be defined in the functional domain, where the fundamental purpose of the system—its reason for existence and the problem it is intended to solve—is explicitly established. Nam P. Suh therefore offers the following unified definition:

\textbf{Complexity} is the measure of uncertainty in achieving a given functional requirement (i.e. “what we want to achieve”) or, equivalently, in understanding system behavior (i.e. “what we want to know”) to within a specified accuracy~\cite[pp.~4]{Suh-2005}. In fact, when the defined goals, expressed as a set of functional requirements (FRs), are effectively met using selected design parameters (DPs), meaning that the \textit{system range} (represented by the probability density function of the selected DP's performance) falls entirely within the \textit{design range} (nominal value ± tolerance), the task is not considered complex~\cite{Suh-2005}.

\begin{figure}[ht]
    \centering
    \includegraphics[width=1.0\linewidth]{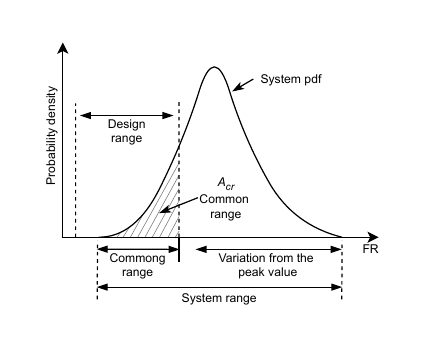}
    \caption{When \textit{system range} is not fully inside the \textit{design range}, we cannot satisfy the functional requirements at all times, therefore, we have complexity.}
    \label{fig:complexity-main}
\end{figure}

\begin{figure}[ht]
    \centering
    \subfloat[Zero\label{fig:complexity-zero}]{%
        \includegraphics[width=0.40\linewidth]{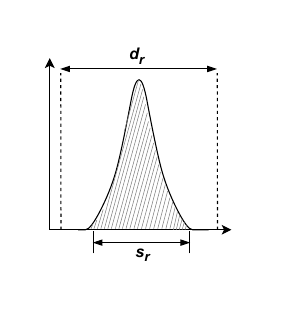}
    }
    \hfill
    \subfloat[Finite\label{fig:complexity-finite}]{%
        \includegraphics[width=0.40\linewidth]{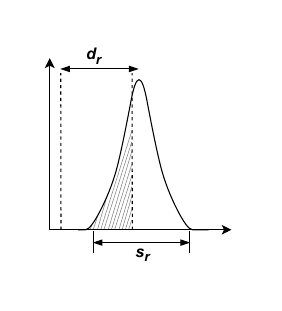}
    }
    \hfill
    \subfloat[Infinite\label{fig:complexity-infinite}]{%
        \includegraphics[width=0.40\linewidth]{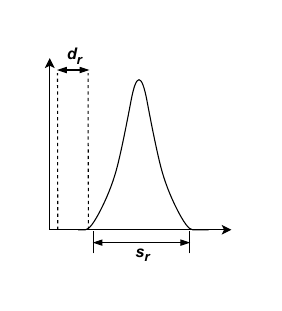}
    }
    \caption{Complexity increases as the design range becomes more restricted.}
    \label{fig:complexity-full}
\end{figure}

For instance, when designing a real-time scheduling task intended to periodically meet a hard deadline, the task remains complex as long as the chosen scheduling algorithm (the design parameter) does not reliably ensure meeting the deadline (the functional requirement) within the specified design range (nominal deadline ± tolerance). However, once the scheduling algorithm is selected and proven—such that the system range consistently lies within the design range—the uncertainty is eliminated, and thus the task is no longer complex. This leads us directly to our adopted definition of complexity for iCAS which we will explore in the following up section.

When multiple FRs are imposed on a system concurrently, the overall complexity of the system is determined by whether or not the DPs chosen to satisfy the FRs couple FRs to each other. The system exhibits minimal complexity when each FR can be satisfied without affecting other FRs. Consequently, in such scenarios, each FR may be fulfilled within its designated range by adjusting its respective DP and by reducing the 'stiffness' of the system~\cite{Suh-2005,Suh-Complexity-2005}.

\section{Formulation Of Complexity}
\label{Formulation_of_complexity}
A design is called complex when its probability of success is low, that is, when the information content (IC) required to satisfy the FRs is high~\cite[pp.~40]{Suh-2001}. Therefore, before we present the formulation of complexity introduced by Suh~\cite{Suh-2005} we must first introduce how information content is being calculated based on axiomatic design~\cite{Suh-2001}.

Information content for $I_i$ for a given $FR_i$ is defined in terms of the probability $P_i$ of satisfying $FR_i$:
\begin{equation}
    I_i = \log_2 \frac{1}{P_i} = -\log_2 P_i
    \label{eq:entropy}
\end{equation}

The information is given in units of bits and the logarithmic function is chosen so that the information content will be additive when there are many FRs that must be satisfied simultaneously. In the general case for \textit{m} FRs, the information content for the entire system $I_sys$ is:

\begin{equation}
    I_{sys} = -\log_2 P_m
    \label{eq:general_entropy}
\end{equation}

The probability of success can be computed by specifying the \textit{design range (dr)} for the FR and by determining the \textit{system range (sr)} that the proposed design can provide to satisfy the FR. Fig.~\ref{fig:complexity-full} represents these two ranges graphically. i.e. when the \textit{sr} and the \textit{dr} have no overlap, the resulting information content becomes infinite due to a zero denominator and when the \textit{sr} entirely falls under the \textit{dr}, then the \textit{common range ($A_{cr}$)} is identical to the \textit{sr}, leading to zero information content. Therefore, minimizing the information content is preferable~\cite{Igata-1996} as it leads to lower complexity.

In principle, the probability of success for a \textit{dr} with lower tolerance is less than a \textit{dr} with higher tolerance, as the task with lower tolerance of error is more complex compared to a task with higher error tolerance. For instance, consider the scenario wherein Rod A must be cut to a length of 1 ± 0.000001 m (±1 µm) versus Rod B, which must be cut to 1 ± 0.1 m. Which scenario possesses a higher probability of success? Furthermore, how would the probability of success be affected if the nominal length of the rod is 30 m instead of 1 m? The answer depends on the cutting equipment available for the task. Nonetheless, most engineers with practical experience would assert that the task requiring precision within 1 µm is more challenging. This is because tasks with smaller tolerances generally correlate with a lower probability of success, reflecting increased complexity. The complexity is then evaluated based on how tightly the tolerances are set around this nominal value, affecting the probability of successfully cutting the rod within the defined tolerance. The smaller the tolerance relative to the nominal length, the more difficult and thus complex the task becomes~\cite{Suh-2005}.

Let us explain the above concept with an example from a real-time embedded system. Consider a real-time embedded system that must periodically execute two independent tasks:
\begin{itemize}
    \item $FR_1$: A sensor monitoring task must sample data precisely every 1 millisecond (± 0.1 µs) without delay.
    \item $FR_2$: A communication task must transmit data packets exactly every 5 milliseconds (± 0.5 µs).
\end{itemize}

Here, we assume that our design parameter (DP) is the scheduling algorithm chosen (e.g., fixed-priority preemptive scheduling or Earliest Deadline First scheduling), with design ranges:
\begin{itemize}
    \item Task 1: Nominal = 1 ms ± 0.1 µs
    \item Task 2: Nominal = 5 ms ± 0.5 µs
\end{itemize}

While system range is the actual task execution intervals achieved by the selected scheduling algorithm, both tasks are presenting a similar challenge and complexity arises if the chosen scheduling algorithm’s timing distribution (system range) cannot consistently meet both $FR_1$ and $FR_2$ within the specified design ranges. If the algorithm consistently achieves timing within these ranges, complexity is low or zero. However, if task executions overlap or miss deadlines due to inadequate scheduling, complexity becomes finite or infinite (task failure).

In Fig~\ref{fig:complexity-full}, the system probability density function (pdf) is plotted over the system range for the specified FR. The intersection of the design range with the system range is identified as the \textit{common range (cr)}, wherein the FR is fulfilled exclusively. Therefore, the integral of the system pdf over this shared range, denoted as $A_{cr}$, represents the probability that the design will meet the specified objective. Subsequently, the information content can be articulated as follows~\cite{Suh-1990}:

\begin{equation}
    I = \log_2 \frac{1}{A_{cr}}
    \label{eq:entropy_common_area}
\end{equation}

In terms of the system pdf $P_s$($FR_i$), the probability of $P_i$ of satisfying $FR_i$ is the integral of $P_s$($FR_i$) over the $FR_i$ design range, which may be expressed as

\begin{equation}
p_i = \int_{\text{design range}} p_s(FR_i)\, dFR_i
\end{equation}

\section{Complexity in Software Systems}
\label{complexity_in_software_systems}
Generally speaking, defining complexity in the field of software engineering falls under the same ambiguity discussed above. Many works either misrepresent the concept or fail to establish a common definition. For example, when Tompkins et al.~\cite{Tompkins-1950} attempted to manage the complexity of problems by dividing them into smaller subproblems in order to reduce the need for computational resources back in 1950, they basically tried to deal with the difficulty of running large tasks on limited resources, i.e., complexity was defined in relation to the constraints imposed by limited computational resources (physical domain).

Similarly, when E. W. Dijkstra discussed the use of 'divide et impera' (divide and rule) to manage program complexity back in 1965~\cite{Dijkstra-1965}, he was essentially referring to the difficulty of solving large problems compared to breaking them into independent subproblems, which were constrained by human cognitive limitations~\cite{Miller-1956}.

In 1986, in the paper "No Silver Bullet," Frederick P. Brooks defines complexity as an essential property of software, stemming from its abstract, interdependent nature and vast number of possible states. He argues that, unlike physical constructs, software lacks repetitive elements, which makes it uniquely intricate for its size. In his view, the complexity of software grows nonlinearly as systems scale, amplifying the challenges of communication, testing, and reliability, resulting in a higher effort for developers to learn the system's behavior. In Brook's belief, this inherent complexity is not accidental but is rooted in the very essence of software systems. Although accidental complexities, such as inefficiencies in tools or processes, can be reduced, the essential complexity of software remains irreducible, fundamentally shaping the challenges of software development~\cite{Brooks-1986}.

Brooks's perception of complexity as an inherent characteristic of software systems stems from his incorrect approach of locating complexity within the physical structure of systems. This led him to conclude that building a software system is inherently a complex task~\cite{Brooks-1972,Brooks-1986}. In contrast, Suh emphasizes that complexity is fundamentally about uncertainty in satisfying the what (functional requirements). Although Suh recognizes that complexity might sometimes correlate with the number of system parts or vary over time, he argues that these attributes alone do not define complexity. Indeed, a system comprising many components could have zero complexity if all functional requirements are consistently met, whereas a simpler system with fewer components could still exhibit significant complexity due to uncertainty in achieving its defined functions. This could be easily proved in the field of software engineering by tasking a software engineer to design a small function to fulfill a functional requirement with an extremely tight tolerance ratio (less than $10^-6$), such as consistently meeting a hard deadline in real-time systems, which could be significantly more complex than building a comparatively large-scale application like a hotel reservation tool.

This is not only a phenomenon of the early days of the software industry but continues to be observed even in recent times. According to ISO/IEC/IEEE 24765:2017, software complexity refers to the difficulty in understanding a system due to numerous components or their interactions, relates to structure-based metrics that measure these characteristics, and indicates the extent to which the design or implementation complicates understanding and verification~\cite[pp.~81]{ISO-IEC-IEEE-24765-2017}.

Just in particular to iCAS, numerous recent instances can be found where scholars and practitioners have investigated complexity within the physical domain (e.g. machines, lines of code, computation time, memory space, size of network of FBs\footnote{Refering to the number of interconnected Function Blocks in graphical languages such as IEC 61499, analogous to lines of source code in text-based programming languages.})~\cite{Hopkinson-1998,Bonfatti-Daniela-2001,Thieme-Hanisch-2002,Katzke-Fischer-Vogel-2004,Sunder-Zoitl-2006,Zoitl-Lewis-2014,Zhabelova-Vyatkin-2015,Alkan-2017,Muslija-2017,Bougouffa-2018,Hag-2019,Sonnleithner-2021,Zhou-2022}, or have employed the term 'complexity' in an intuitive manner, presuming that the reader possesses an implicit understanding of its definition, which is largely unreasonable~\cite{Thramboulidis-2012,Vogel-Heuser-2015,Pakonen-Buzhinsky-Vyatkin-2024,Ovsiannikova-Vyatkin-2024}. The reason why we see many different understandings and definitions of complexity is that researchers and practitioners often approach it in the physical domain, influenced by their specific domains, methodologies, and underlying assumptions.

The only study prior to our study that tried to present a common definition to complexity in the field of iCAS is the one worked by Alkan et al.~\cite{Alkan-2017, Alkan-2019}. They attempt to leverage Nam P. Suh's definition of complexity to provide a common definition for 'complexity' in the field of iCAS; however, the interpretation appears incomplete, as they exclusively refer to an early version of Suh's work~\cite{Suh-1995}, which addresses complexity only briefly within the context of axiomatic design. Consequently, their focus is limited solely to information content, formulating it as a function of modular dependencies, an approach that is fundamentally flawed due to being defined in the physical domain~\cite{Suh-2005}.

In an iCAS, complexity could be defined as the degree of uncertainty in satisfying effectively the control domain needs. The \textit{design range} is defined in the control domain and the \textit{system range} is the performance provided by the automation domain for a given process that has to be automated in the control domain. A solution is considered effective if it has the lowest complexity, i.e. complexity is zero.

\begin{figure}
    \centering
    \includegraphics[width=1\linewidth]{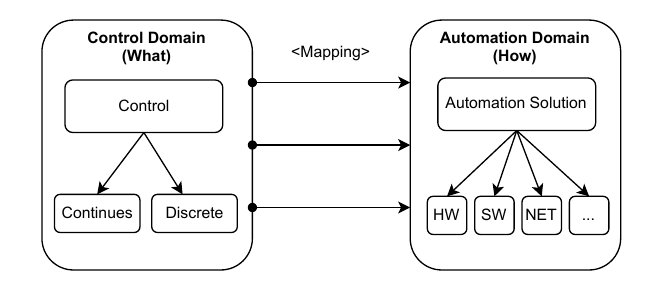}
    \caption{Design domains in iCAS.}
    \label{fig:icas_design}
\end{figure}

\section{What is Coupling}
\label{what_is_coupling}
The problem of coupling that causes complexity in software systems was observed well before the principle itself was established. Dijkstra wrote in 1968 a classic paper against the "goto" statement, pointing out that the transfer of control from one module to another module has disastrous effects~\cite{Dijkstra-1968a}. Later, this was called "\textit{Control Coupling}"~\cite{Myers-1975,Yourdon-1978,DeMarco-1979,Page-Jones-1988} which caused the software system to potentially behave unpredictably when a new change was wanted to be introduced as part of system expansion or correction.

Traditionally, in software systems, coupling is defined by Myers (in 1975)~\cite{Myers-1975}, Yourdon and Constantine (in 1978)~\cite{Yourdon-1978}, DeMarco (in 1979)~\cite{DeMarco-1979}, and Page-Jones (in 1988)~\cite{Page-Jones-1988} as the degree of intermodular relatedness (that is, the relationship among modules). They further differentiate between \textit{highly coupled}, \textit{loosely coupled}, and \textit{decoupled} intermodular relatedness, indicating that highly coupled modules are strongly interconnected, loosely coupled modules have weak interconnections, and uncoupled or decoupled modules have no interconnections. They argue that if two modules are interconnected, a modification in one module will necessitate alterations in the interconnected module, and the effort of implementing this change will determine its strength.

Empirical studies show that the spectrum of coupling (loose-tight) has since been widely accepted by various others~\cite{ Lethbridge-2005,Page-Jones-1980,Gamma-1994,Pressman-2015,Newman-2015,Kent-2023}, although with slight variations. Therefore, more recent definitions of coupling do not differ much from those of other studies during the 1970s and 1980s. For example, Grady Booch defines coupling in object-oriented design as the measure of the strength of association established by a connection from one module to another~\cite{Booch-2007}, or Kent Beck defines coupling between two modules as the result of change cascading from one module to another module~\cite{Kent-2023}.

The Table~\ref{tab:structured_coupling} illustrates the coupling spectrum as defined in Structured Design~\cite{Yourdon-1978}, the Table \ref{tab:oop_coupling} presents the extended version of the same spectrum within Object-Oriented Programming~\cite{Lethbridge-2005} .

\begin{table}[h]
    \caption{Coupling in Structured Design}
    \label{tab:structured_coupling}
    \centering
    \begin{tabular}{|c|p{4.5cm}|c|}\hline
        \textbf{Type} & \textbf{Description}  & \textbf{Spectrum} \\\hline
        Content & Occurs when some or all of the contents of one module are included in the contents of another. & Tight \\\hline
        Common  & A group of modules are common coupled if they reference a shared global data structure (a common environment). & \\\hline
        Hybrid & Occurs when connection to the target (modified) module functions as control and to the modifying module as data. & \\\hline
        Control & Transfer of control (e.g., activation of modules), or passing of data that change, regulate, or synchronize the target module. & \\\hline
        Data & The outputs of some modules become the inputs of others. & Loose\\\hline
    \end{tabular}
\end{table}

\begin{table}[h]
    \caption{Coupling in Object-Oriented Design}
    \label{tab:oop_coupling}
    \centering
    \begin{tabular}{|c|p{4cm}|c|}\hline
        \textbf{Type} & \textbf{Description}  & \textbf{Spectrum} \\\hline
        Content & A component is trying to spoof the internal data of another component. & Tight \\\hline
        Common  & The use of global variables.  & \\\hline
        Control & One procedure directly controls another using a flag. & \\\hline
        Stamp & One of the argument types of a method is one of your application classes. & \\\hline
        Data & The use of method arguments that are simple data. & \\\hline
        Routine call & A routine calling another. & \\\hline
        Type use & The use of a globally defined data type.  & \\\hline
        Inclusion/import & Including a file or importing a package. & \\\hline
        External & A dependency exists on elements outside the scope of the system, such as the operating system, shared libraries, or hardware. & Loose \\\hline
    \end{tabular}
\end{table}

Larman~\cite{Larman-2005} and Lethbridge~\cite{Lethbridge-2005}, state that coupling occurs when one module depends on another. They identify three key indicators of interdependency: (1) \textit{rippling change}, where modifications in one module affect all dependent modules, (2) \textit{difficulty in understanding} the functionality of a coupled module, and (3) increased effort in \textit{reusability} due to inherent physical dependencies.

The issue with all of the above studies is their attempt to define \textit{coupling} within the physical world, where a high level of interconnectivity is prevalent, resulting in the erroneous belief that it is infeasible to effectively address coupling in software systems~\cite[pp.~308]{DeMarco-1979}. Consequently, the focus should be directed towards minimizing it, moving away from \textit{tight coupling} to \textit{loose coupling}. In fact, this inadequate definition of optimal design as "high cohesion and low coupling," which has persisted since 1970~\cite[pp.~33]{Myers-1975} to the present day~\cite[pp.~38-41]{Newman-2021}, arises from the notion that designing decoupled or uncoupled software is unachievable.

From Nam P Suh's research in the domain of design and complexity~\cite{Suh-1990,Suh-2000,Suh-2005} is evident that like \textit{complexity}, \textit{coupling} should also be defined within the functional domain rather than the physical domain. A design is considered coupled when modifying a design decision~\footnote{Design decisions, or more specifically, design elements, serve as embodiments of solutions intended to fulfill functional requirements. Suh calls them Design Parameters~\cite[pp.~10]{Suh-2000}.} affects the satisfaction of multiple functional requirements, thereby violating FRs independence. However, if altering one design element influences another design element without impacting other independent functional requirements, then the design remains inherently decoupled and no further decoupling is required~\cite[pp.16–19]{Suh-2000}. This definition of coupling aligns closely with the Single Responsibility Principle (SRP) discussed by Robert C. Martin\cite{R.C.Martin-2000}. Of course, this does not mean that we should neglect interface optimization. Minimizing the number of interfaces between design elements and restricting cascading changes across interfaces remains essential but cannot necessarily present coupling.

\begin{figure}[h]
    \centering
    \subfloat[Coupled\label{fig:coupled}]{%
        \includegraphics[width=0.45\linewidth]{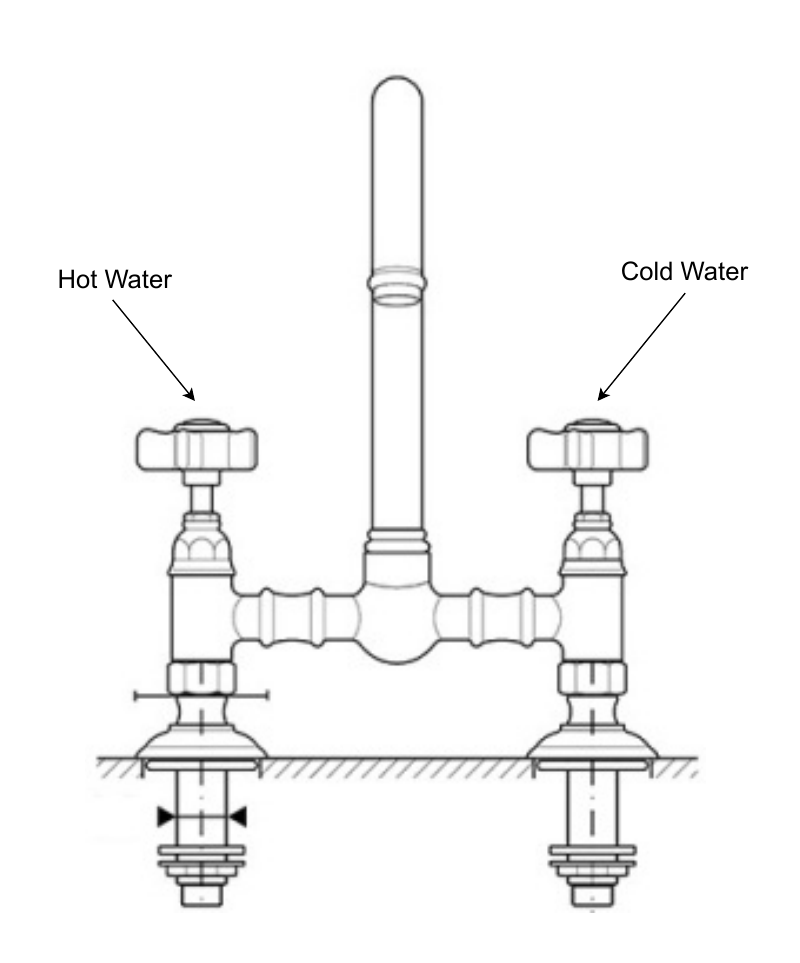}
    }
    \hfill
    \subfloat[Uncoupled\label{fig:uncoupled}]{%
        \includegraphics[width=0.40\linewidth]{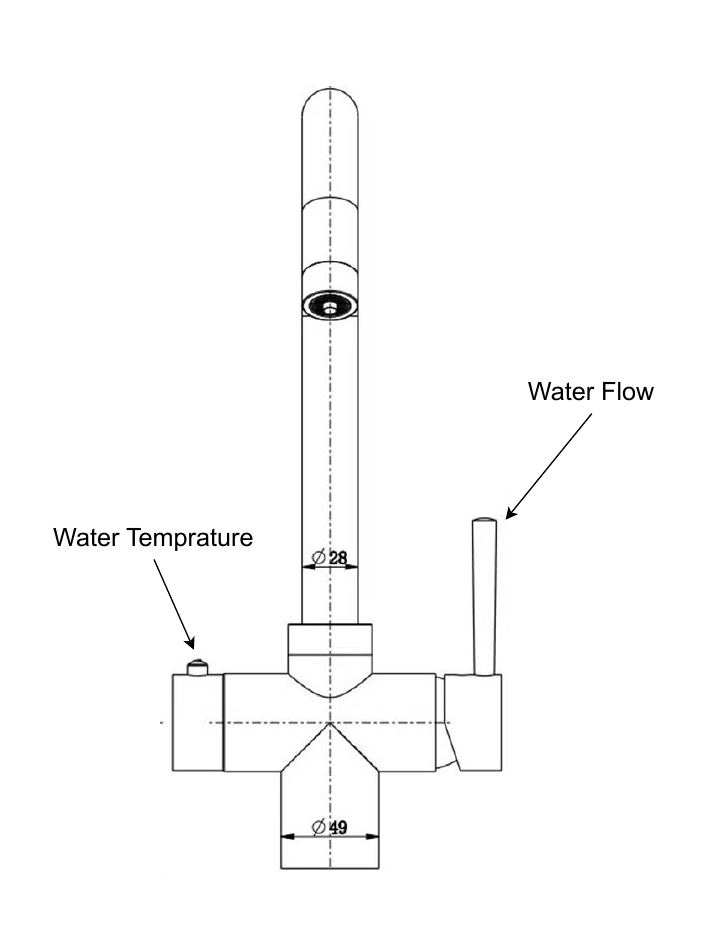}
    }
    \caption{The figure on the left represents a coupled design, while the one on the right represents an uncoupled design.}
    \label{fig:coupled_uncoupled_faucet}
\end{figure}

We shall demonstrate this by employing the analogy of a kitchen faucet. In this instance, a survey was administered involving over 100 software engineers, during which Figure~\ref{fig:coupled_uncoupled_faucet} was presented to them. The participants were assigned the task of distinguishing between coupled and uncoupled designs in relation to the specified functional requirements:

\begin{itemize}
    \item $FR_1$: control the flow ration of the water.
    \item $FR_2$: control the temperature of water.
\end{itemize}

Although the figure on the left represents a coupled design, since the cold and hot water knobs affect both functional requirements, more than 80\% of the participants incorrectly identified the left figure (\ref{fig:coupled}) as an uncoupled design.

If we consider coupling as an interconnection between two knobs, such that changing physical elements in one knob requires changes in the physical elements of another (the traditional definition of coupling in software), then the figure on the left would indeed be decoupled, because modifying one knob would not necessitate altering the other. For example, if the cold water knob needs to be repaired or replaced, the hot water knob remains unaffected. However, within the functional domain, the situation is quite the opposite. The faucet on the right is functionally uncoupled because, at any given moment, each design parameter independently controls either the water flow rate or its temperature. And this functional clarity is evident in practice—virtually anyone who has used a modern faucet even once can attest that it provides a significantly easier and more convenient experience compared to the classic design on the left.

By extending this analogy to software and employing proof by contradiction, let us examine two services, modules, components, classes, functions or even function blocks in the case of IEC 61131-3 and IEC 61499 that change due to the same reason. Can coupling then be defined simply as a cascading change, even though both changes address the same independent requirement? Clearly, the answer is no, unless each change impacts more than one independent requirement. 

This is exactly what the Single Responsibility Principle (SRP) at the class level and the Common Closure Principle (CCP) at the component level tries to convey~\cite{R.C.Martin-2000}. The SRP emphasizes that each class should have exactly one reason to change, which means that it should only address a single independent requirement. Similarly, the CCP extends this notion to higher-level software components, advocating that components should encapsulate functionalities that typically change together for the same reasons\footnote{In the context of axiomatic design, when two functional requirements are not independent, they are combined into a single functional requirement~\cite{Suh-2000}.}.

A similar idea has been investigated by Homay et al.~\cite{Homay-Sousa-2020,Homay-Wollschlaeger-2024} through the concept of Process Granularity (PG). PG aims to define the appropriate granularity of each Function Block (FB)\footnote{In IEC 61131, a Function Block (FB) is a reusable software component (i.e., program orgnization unit) that encapsulates control logic and maintains internal state and data between executions~\cite{TC65/61131-3:2013}.}. This is accomplished by first identifying independent control processes (i.e. control temperature, control level, and control mixer)~\cite{Homay-Wollschlaeger-2024} and then assigning an appropriate granularity to the Program Organization Unit (e.g., Program, Function Block~\cite{TC65/61131-3:2013}) so that each entire process is encapsulated within a single element. In essence, PG helps ensure that each element (e.g., FB, Program) has a clear boundary and focuses exclusively on one independent requirement, thus avoiding coupling and reducing complexity.

The independence axiom within the framework of axiomatic design offers a more comprehensive and appealing characterization of coupling. This axiom states that a design should be structured so that each functional requirement is satisfied without affecting other functional requirements (uncoupled). The goal is to ensure that the design parameters (DPs) chosen to achieve one FR do not inadvertently affect other FRs. In other words, each FR should be controlled by one set of DPs without causing unintended interactions with other FRs. After the FRs are established, conceptualization is the next step in the design process. This occurs during mapping, going from the functional (what) domain to the physical (how) domain. Recall that during the mapping process, we must think of different ways to satisfy each of the FRs by identifying plausible DPs to fulfill the independence axiom. The following mathematical model represents the mapping process (design equation):
\begin{equation}
    \{FR\} = [A]\{DP\} \label{eq:calculate_FRs}
\end{equation}

Where \textit{[A]} is a matrix (called a design matrix) that relates FRs to DPs. The~\ref{eq:design_matrix_A} represents the design matrix. The independence axiom will be satisfied if the design matrix is diagonal or triangular.

\begin{equation}
    A = \begin{bmatrix}
        A_{11} & A_{12} & A_{13} \\
        A_{21} & A_{22} & A_{23} \\
        A_{31} & A_{32} & A_{33}
    \end{bmatrix}
    \label{eq:design_matrix_A}
\end{equation}

When Eq.~\ref{eq:design_matrix_A} is written in differential form as:
\begin{equation}
    \{dFR\} = [A]\{dDP\} \label{eq:calculate_FRs_differential}
\end{equation}

The elements of the design matrix with three FRs and DPs are given by

\begin{equation}
    A_{ij} = \frac{\partial FR_i}{\partial DP_j} \label{eq:placeholder}
\end{equation}

Eq.~\ref{eq:calculate_FRs} can be written in a general form as:

\begin{equation}
    FR_{i} = \sum_{j=1}^{n} A_{ij} \cdot DP_{j}
    \label{eq:design_matrix_A_general}
\end{equation}

If the resulting design matrix is diagonal, then the design is \textit{uncoupled}, which means that each of the FRs can be satisfied independently employing its respective DP. However, in the case of a triangular design matrix (\textit{decoupled}), the independence of FRs can only be guaranteed if and only if the DPs are determined in a proper sequence. Otherwise, we have a complete matrix, resulting in a \textit{coupled} design (cf.~\ref{fig:design_matrices}).

\textbf{Uncoupled Design} In an uncoupled design, each Functional Requirement (FR) is ideally controlled by a single Design Parameter (DP). This creates a diagonal design matrix, indicating that modifications in a DP only impact their associated FR.

\textbf{Decoupled Design} In a decoupled design, the design matrix is triangular, which means that some DPs may influence multiple FRs, but the dependencies are structured in a manageable sequence. Conflicts among FRs and DPs in designing large systems are not preventable. FRs often form a conflict, making the selection of DPs very difficult. An approach to solving potential conflicts is to use decomposition~\cite{Suh-2005}.

\textbf{Coupled Design}
If a design matrix is not uncoupled (diagonal) or decoupled (triangular), it is a coupled design matrix that leads to a coupled system.

\begin{figure}[ht]
\centering
\[
\setlength{\arraycolsep}{1pt} 
\begin{array}{ccc}
\begin{bmatrix}
A_{11} & 0 & 0 \\
0 & A_{22} & 0 \\
0 & 0 & A_{33}
\end{bmatrix}
& \quad
\begin{bmatrix}
A_{11} & 0 & 0 \\
A_{21} & A_{22} & 0 \\
A_{31} & A_{32} & A_{33}
\end{bmatrix}
& \quad
\begin{bmatrix}
A_{11} & A_{12} & A_{13} \\
A_{21} & A_{22} & A_{23} \\
A_{31} & A_{32} & A_{33}
\end{bmatrix}
\\
(a) & (b) & (c)
\end{array}
\]
\caption{The (a) represents uncoupled, (b) represents decoupled, and (c) represents coupled design matrices.}
\label{fig:design_matrices}
\end{figure}

\section{Running Example}
\label{running_example}
For the purpose of this example, we elected to select the control and automation domain, as it facilitates a more straightforward representation of both the tangible and intangible components of software systems compared to other domains.

Fig~\ref{fig:running_example} represents a continuous process control relevant to a milk processing production facility. This ongoing process is engineered to thermally condition and amalgamate (mix) milk prior to its bottling. The procedure initiates with the assessment of the tank's fluid level while the outlet valve remains open. Should the liquid level inside the tank descend below one meter, the outlet valve is to be closed and the inlet valve opened. This sequence persists until the fluid ascends to a level of 7 meters. Upon attaining this level, the inlet valve is closed, and the temperature of the fluid is measured, highlighting the interaction between level and temperature parameters. In instances where the temperature is under 65 degrees Celsius, the heater is activated to elevate the temperature to precisely 65 degrees Celsius, after which the heater is deactivated. The mixer then operates for a duration of six seconds to ensure thorough mixing, followed by its deactivation and the reopening of the outlet valve. Conversely, if the temperature is already 65 degrees Celsius when the fluid level reaches 7 meters, the heater is not engaged. The mixing apparatus functions for six seconds, after which the outlet valve is reopened, underscoring the linkage between temperature and mixing operations. The practical scenario incorporates two transmitters: one assigned to gauging the tank level, measuring a range from 0 to 8 meters, with the address PIW802, and the other designated for temperature measurement, covering a span from 0 to 100 degrees Celsius, possessing the address PIW800. The high level functional requirements are defined as follows:

\begin{figure}[ht]
    \centering
    \includegraphics[width=0.8\linewidth]{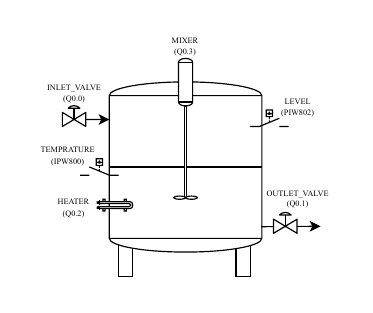}
    \caption{Tank for heating and mixing milk before releasing to bottle filler.}
    \label{fig:running_example}
\end{figure}

For this running example, the functional requirements (FRs) are defined as follows:

\begin{itemize}
    \item \textit{FR\textsubscript{1}}: Maintain the milk tank level below 7 meters, without affecting the temperature.
    \item \textit{FR\textsubscript{2}}: Maintain the temperature at exactly 65°C, without affecting the level.
    \item \textit{FR\textsubscript{3}}: Mix milk effectively before filling bottles, without influencing temperature or level.
\end{itemize}

The corresponding design parameters (DPs) are implemented as separate software-based PID controllers or logic blocks written in compliance with the IEC 61131-3 standard:

\begin{itemize}
    \item \textit{DP\textsubscript{1} (Level\_Control FB)}: Controls inlet and outlet valves to regulate milk level.
    \item \textit{DP\textsubscript{2} (Temp\_Control FB)}: Regulates heating/cooling mechanisms to maintain the required temperature.
    \item \textit{DP\textsubscript{3} (Mixer\_Control FB)}: Activates the mixer mechanism independently of other processes.
\end{itemize}

\textbf{Design Ranges (tolerances):}
\begin{itemize}
    \item Level: 7 meters ± 5 cm
    \item Temperature: 65°C ± 0.5°C
    \item Mixing: Nominal mixing cycle of 120 seconds ± 2 seconds
\end{itemize}

\textbf{System Range:} The actual values achieved for level, temperature, and mixing duration during operation, represented by their corresponding probability density functions (pdfs).

\textbf{Complexity and Coupling Analysis:}
\begin{itemize}
    \item If \textit{DP\textsubscript{1}} (Level\_Control) inadvertently affects temperature or mixing due to shared physical components (e.g., a pump used for both draining and mixing), the design is functionally \textit{coupled}.
    \item If \textit{DP\textsubscript{2}} (Temp\_Control) unintentionally impacts the level or mixing—such as through thermal expansion or convection—coupling occurs.
    \item If \textit{DP\textsubscript{3}} (Mixer\_Control) indirectly influences the level or temperature (e.g., due to turbulence or heat dispersion caused by mixing), the design again becomes coupled.
\end{itemize}

\begin{figure*}[ht]
    \centering
    \includegraphics[width=1\linewidth]{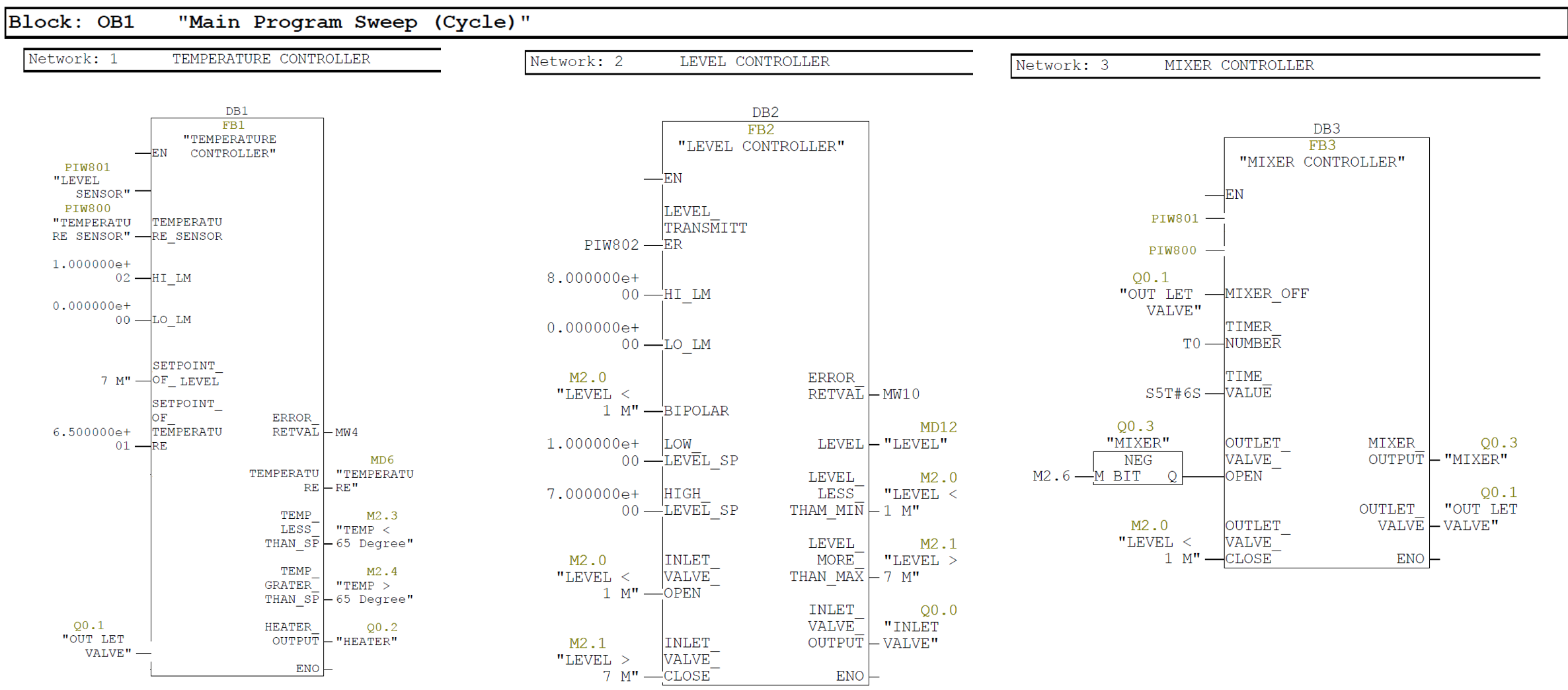}
    \caption{In PG (Process Granularity) each control process has been defined as an independent functionl requirements and seperated in an independent function block~\cite{Homay-Wollschlaeger-2024}.}
    \label{fig:running_example_imp}
\end{figure*}

The ideal design is an \textit{uncoupled design}, where each function block (FB) satisfies exactly one independent FR without influencing the others. Any coupling increases the system’s complexity because it raises the uncertainty of consistently achieving all FRs within their respective design ranges. This idea design is implemented in STEP7 (Siemens control system design tool) and has been represented in Fig.~\ref{fig:running_example_imp}.

\section{Why Coupled Design is Complex}
\label{why_compled_design_is_complex}
In an uncoupled design like the one presented in the running example, the information content for the entire system is calculated by Eq.~\ref{eq:general_entropy} where the $P_m$ is the joint probability that all \textit{m} FRs are satisfied. When all FRs are statistically independent, as is the case for an uncoupled design, 

\begin{equation}
P_{\{m\}} = \prod_{i=1}^{m} P_i
\end{equation}

The $I_{sys}$ may be expressed as

\begin{equation}
I_{\text{sys}} = \sum_{i=1}^{m} I_i = -\sum_{i=1}^{m} \log_2 P_i
\end{equation}

When all FRS are not statistically independent, as is the case for a decoupled design, 

\begin{equation}
P_{[m]} = \prod_{i=1}^{m} P_{i|(j)} \quad \text{for } j = \{1, 2, \ldots, i-1\}
\end{equation}

\noindent
where \( P_{i|(j)} \) is the conditional probability of satisfying \( FR_i \), given that all previously defined relevant \( FR_j \) (with \( j < i \)) are already satisfied. In this case $I_{sys}$ may be expressed as

\begin{equation}
I_{\text{sys}} = -\sum_{i=1}^{m} \log_2 P_{i|(j)} \quad \text{where } j = \{1, 2, \ldots, i-1\}
\label{eq:P_decoupled}
\end{equation}

In a coupled design, the FRs are \textbf{not independent}, which means the design violates the Independence Axiom~\cite[pp.~16]{Suh-2001}. In such designs, a single design parameter (DP) affects multiple FRs, and the design matrix contains significant off-diagonal terms. As a result, the probability of satisfying each FR cannot be computed sequentially using conditional probabilities because the satisfaction of any one FR inherently depends on the values of other FRs. Even though such a design (coupled design) is not acceptable in axiomatic design, one may try to calculate the total system probability from the \textbf{full joint probability distribution} \( p_s(FR_1, FR_2, \ldots, FR_m) \), integrated over the \textbf{multidimensional design space}:

\begin{align}
P_{1,2,\ldots,n} = \int_{\text{design space}} & \, p_s(FR_1, FR_2, \ldots, FR_m) \nonumber \\
& \times\, dFR_1, dFR_2, \ldots, dFR_m
\label{eq:p_coupled}
\end{align}

As it is illustrated from Eq.~\ref{eq:P_decoupled} and Eq.~\ref{eq:p_coupled} coupled design is complex because it introduces considerable uncertainty into the design process~\cite{Suh-2005}. This complexity arises from the fundamental nature of coupled relationships, where one design parameter~\footnote{A design parameter (DP) is a physical, structural, or conceptual solution chosen to fulfill a specific functional requirement~\cite{Suh-2000}.} simultaneously affects multiple functional requirements (FRs)~\footnote{Functional requirments are indepdent by definition~\cite{Suh-2000}.}. In such a scenario, adjusting one parameter to optimize or satisfy a particular requirement invariably influences other requirements, often unpredictably, leading to intricate dependencies and challenging compromises.

According to Suh, in a coupled design, the design matrix—representing the relationships between FRs and DPs, is neither diagonal nor triangular, meaning each parameter influences multiple functions in a nonlinear and interdependent manner. Such interdependencies lead to complexities because a designer cannot simply adjust one parameter in isolation without inadvertently creating unintended changes or conflicts in other areas of the design. This characteristic significantly reduces predictability and control within the design process, amplifying uncertainty~\cite{Suh-2000}.

Furthermore, Herbert Simon emphasizes the role of hierarchical complexity in systems, explaining that complex systems often consist of layers of subsystems interconnected through various relationships. When these subsystems are coupled, it becomes difficult to predict their collective behavior, as changes propagate throughout the system in unexpected ways~\cite[pp.~183-187]{Simon-1996}. Newell and Simon (1972) also highlighted that human problem-solving capabilities are inherently limited when managing interdependent information, further complicating the design process as cognitive boundaries are reached quickly in coupled designs~\cite[pp.~872-873]{Newell-Simon-1972}.

Thus, coupled design inherently involves a continuous balancing act, where the designer must navigate intricate networks of interactions between parameters and requirements. This intricate web makes it challenging to achieve stable, robust solutions, necessitating extensive iterative adjustments and compromises. Consequently, the uncertainty inherent in coupled designs not only complicates the design process but also substantially increases the risk of overlooking critical interactions, potentially leading to failures, suboptimal performance, and increased resource expenditures in terms of time, cost, and cognitive effort.

In a coupled design, optimizing the system range by choosing a better design parameter to reduce or eliminate complexity for one functional requirement (FR) can inadvertently influence the system range of another FR, shifting its complexity from zero complexity to finite or even infinite complexity. Additionally, in such coupled design, the design ranges often become narrower because multiple FRs must be satisfied simultaneously, significantly decreasing the probability of successfully meeting each FR within its specified design range. Therefore, an effective design has to be uncoupled or decoupled.

\section{Conclusion}
\label{conclusion}
This paper has provided clear, scientific definitions of complexity and coupling within industrial control and automation systems by employing Nam P. Suh’s groundbreaking research in the field of design and complexity. Complexity is fundamentally characterized by uncertainty in meeting functional requirements defined in the automation domain, while coupling reflects interdependencies among FRs due to shared design parameters (FBs). Our analysis demonstrates that coupled designs inherently increase complexity by restricting the allowable range for satisfying FRs, thus heightening uncertainty and complicating design optimization. Through illustrative examples from different fields including the field of industrial automation, we have shown that, first, coupling and complexity both occur in the functional domain, not in the physical domain, and second, uncoupled or decoupled systems significantly reduce complexity, enhance predictability, and facilitate robust design. Consequently, designers and engineers should aim to systematically minimize coupling to achieve low-complexity systems. Future research will further explore a redefinition of the problem statement that has to be addressed in the field of industrial control and automation systems in terms of its complexity and design quality and will formulate how to understand coupling and complexity to better advance the related field.

\section*{Acknowledgment}
I would like to express my sincere gratitude to Professor Nam P. Suh for his invaluable support and guidance, which significantly deepened my understanding of coupling and the complexity of coupled design.

\bibliography{references}
\bibliographystyle{IEEEtran}

\vfill

\end{document}